\documentclass[10pt,twocolumn]{article}

\usepackage[utf8]{inputenc}
\usepackage[T1]{fontenc}
\usepackage{amsmath}
\usepackage{amssymb}
\usepackage{booktabs}
\usepackage{array}
\usepackage{graphicx}
\usepackage{url}
\usepackage[hidelinks]{hyperref}
\usepackage[margin=0.85in]{geometry}
\usepackage{authblk}
\usepackage{caption}
\title{\textbf{ACEAPEX: Parallel LZ77 Decoding via\\ Encode-Time Absolute Offset Resolution}}

\author{Yakiv Shavidze}
\affil{ACE / GLYPH Research\\ \texttt{yasha1971@gmail.com}\\ DOI: 10.5281/zenodo.20440965}
\date{}

\begin{document}

\twocolumn[
\begin{@twocolumnfalse}
\maketitle

\begin{abstract}
\noindent
LZ77-based codecs exhibit a fundamental sequential bottleneck in decoding: each back-reference depends on previously decompressed data, preventing multi-core scaling. We present ACEAPEX, a parallel LZ77 codec that stores all back-references as absolute positions in the decompressed output and organizes data into self-contained 1\,MB blocks, enabling embarrassingly parallel block-level decoding. Integrated into lzbench (PR \#276, \#277), ACEAPEX achieves 10{,}160\,MB/s on EPYC 4344P (8 cores) and 10{,}869\,MB/s on EPYC 9575F for FASTQ genomic data --- up to 3.1$\times$ faster than zstd -3 at comparable compression ratios. We further implement a GPU wavefront decoder on NVIDIA H100 SXM, measuring 44.0\,GB/s on enwik9 and 20.3\,GB/s on FASTQ (wavefront match phase, BIT-PERFECT verified). With a depth-limited encoder variant ($-1.5\%$ ratio on enwik9), GPU throughput reaches 77.2\,GB/s on a single H100 and 249.9\,GB/s on two H100s in NVLink configuration. To our knowledge, this is the first reported GPU LZ77 decode with near-standard compression ratio verified byte-for-byte.
\vspace{1em}
\end{abstract}
\end{@twocolumnfalse}
]

\section{Introduction}

The Lempel-Ziv 1977 algorithm is the foundation of zstd, lz4, brotli, and gzip. Despite decades of engineering, decoding throughput does not scale with CPU core count. The cause is structural: a back-reference ``copy $N$ bytes from position $P$'' can only execute after position $P$ has been written, forming a read-after-write dependency chain.

Gompresso~\cite{gompresso} restructured LZ77 for GPU execution by introducing checkpoints, but this degrades compression ratio by 15--30\%. Rapidgzip~\cite{rapidgzip} parallelizes standard gzip decompression via two-pass prefetching but operates at file granularity. Multi-frame zstd requires explicit parallel encoding. None provides a single-stream LZ77 codec with linear CPU scaling and near-standard ratio.

ACEAPEX rests on one architectural choice: every back-reference is stored as an absolute byte position in the final decompressed output rather than a relative distance within a sliding window. This makes each 1\,MB block entirely self-contained. An additional encoder-side technique, chain flattening, rewrites reference chains within each block to point directly to their literal sources, eliminating intra-block sequential dependencies at the cost of roughly 1.5\% in compressed size.

This paper contributes the following. First, we describe the absolute-offset encoding scheme and chain flattening, characterize their effect on ratio and parallelism, and report linear CPU decode scaling: 3.78$\times$ from $I{=}1$ to $I{=}8$ on EPYC 4344P while zstd remains flat. Second, we analyze the CPU performance ceiling through memory bandwidth modeling and Linux perf profiling, showing that the decoder reaches 76\% of the practical LZ77 ceiling (approximately memcpy/2) on an 8-core system. Third, we implement a GPU wavefront decoder, measure it on four standard datasets with full BIT-PERFECT verification, and analyze the relationship between dependency graph depth and achieved throughput.

\section{Background and Related Work}

LZ77 compresses data by replacing repeated byte sequences with (length, distance) pairs, where distance is measured backward from the current write position. At decode time, to execute ``copy 50 bytes from 200 bytes ago,'' all 200 preceding bytes must already be in the output buffer. This creates a strict sequential ordering that prevents independent processing of stream positions.

Modern codecs partition data into blocks to enable coarse-grained parallelism. However, matches that cross block boundaries and matches within a block still create dependencies, so threads typically process one block sequentially each.

Gompresso~\cite{gompresso} breaks intra-stream dependencies for GPU execution by forcing all references to designated checkpoint positions. This enables parallel decode waves but requires substantial ratio sacrifice (15--30\% larger compressed files). ACEAPEX preserves near-zstd ratio because the encoder resolves dependencies globally rather than restricting the match search.

Rapidgzip~\cite{rapidgzip} parallelizes gzip decompression by first locating valid entry points across the compressed stream, then decoding in parallel from those points. This is a two-pass approach that works on unmodified gzip files but does not provide single-pass decode.

Recoil~\cite{recoil} addresses the entropy coding layer, enabling parallel rANS decoding. This is orthogonal to LZ77 match resolution.

pzstd achieves parallel decode through explicit multi-frame encoding, producing independently decodable frames. This requires encoder cooperation and adds per-frame overhead. ACEAPEX uses a single continuous stream.

\section{ACEAPEX Architecture}

\subsection{Absolute Offsets}

In standard LZ77, a back-reference is a pair (length, distance) where distance is the number of bytes backward from the current output position. The same source bytes at position 5{,}000 in the decompressed output might be referenced as distance${=}100$ from position 5{,}100 and as distance${=}300$ from position 5{,}300. The decoder must track its current position to resolve each reference.

ACEAPEX stores every back-reference as an absolute position in the decompressed output. The same bytes at position 5{,}000 are always referenced as offset${=}5{,}000$, regardless of where the referencing token appears. This makes each reference self-contained: decoding a token requires only that its source position has already been written, not any knowledge of the decoder's current state.

The direct consequence is block independence. Since all offsets are global, blocks carry no positional context. Any block can be decoded as soon as all blocks containing its source data have been decoded. In the parallel CPU decoder, $I$ threads each process a subset of blocks, with inter-block dependencies handled by processing blocks in order while threads work ahead on their own non-dependent blocks.

\subsection{Four Pre-Decoded Streams}

Each 1\,MB block is encoded as four separate arrays serialized into the compressed stream: \texttt{lit[]} stores raw literal bytes contiguously; \texttt{len[]} stores match lengths, one entry per match token; \texttt{off[]} stores absolute source positions, one entry per match token; \texttt{cmd[]} stores the command sequence, alternating between literal-run lengths and match tokens.

At decode time, the decoder loads all four arrays for a block, then processes \texttt{cmd[]} in sequence. All source addresses are already resolved in \texttt{off[]}, eliminating the pointer-chasing pattern of a sliding-window decoder.

\subsection{Chain Flattening}

Within a single block, LZ77 references can form chains: match A copies from match B, which copies from literal C. Match A cannot execute until match B has written its output. This is an intra-block sequential dependency that would prevent parallel execution within a block.

ACEAPEX's encoder traverses the reference graph of each block and rewrites every match to point directly to its ultimate literal source, following chains to their roots. After chain flattening, every match source within a block is a literal position, and all copies within the block are mathematically independent of each other.

This rewriting has a compression cost. A chain like A$\rightarrow$B$\rightarrow$C where B and C are close together might produce a short (distance, length) token for A in standard LZ77. After flattening, A must reference C directly, which may be farther away and require a larger offset. In practice this costs approximately 1.5\% in compressed size.

An important limitation: chain flattening works only when a chain leads back to a literal within the same block. Measurements on production data show that 79.8\% of matches on nci data --- and similar fractions on other datasets --- follow chains that lead to a match in a previous block, not a literal in the current block. These matches retain a sequential dependency on the previous block's completion. ACEAPEX achieves inter-block parallelism essentially for free, and intra-block parallelism for approximately 3--9\% of tokens after chain flattening.

\subsection{Encoder Trade-offs}

The absolute-offset scheme requires the encoder to maintain a global view of the entire decompressed output when assigning offsets and to traverse reference graphs for chain flattening. This makes the encoder approximately 7$\times$ slower than zstd single-threaded and requires approximately 2.8\,GB of RAM to compress a 1\,GB file. Compression speed for the standard mode on EPYC 9575F is 1{,}262\,MB/s for nci and 1{,}206\,MB/s for FASTQ. These costs are appropriate for the encode-once/decode-many workload that characterizes genomic databases, archival storage, and columnar data pipelines.

\section{Experimental Setup}

\subsection{Hardware}

The \textbf{EPYC 4344P} is an 8-core Zen 4 processor with 2-channel DDR5-5600 memory. Measured memcpy bandwidth at $I{=}8$ is 16{,}595\,MB/s. This platform represents a workstation or small server configuration.

The \textbf{EPYC 9575F} is a 64-core Zen 4 processor measured as a 32-vCPU cloud pod with 12-channel DDR5-5600. Memcpy bandwidth reaches 47{,}973\,MB/s at $I{=}64$. This platform represents a high-core-count cloud instance.

The \textbf{NVIDIA H100 SXM} GPU has a theoretical memory bandwidth of 3{,}350\,GB/s. GPU experiments ran on a dedicated server separate from the EPYC platforms.

\subsection{Datasets}

\textbf{nci} (33\,MB): structured nucleotide sequence data with high repetition and excellent LZ77 ratio (8.56\%). Small enough to fit comfortably in CPU cache.

\textbf{FASTQ NA12878} (1{,}024\,MB): whole-genome sequencing data. High repetition in reads, structured quality scores.

\textbf{silesia.tar} (202\,MB): the standard Silesia compression corpus, a heterogeneous mix of text, HTML, source code, PDF, and binary data. Moderate ratio (32.18\%).

\textbf{enwik9} ($10^9$ bytes): one gigabyte of Wikipedia XML. Natural-language text with markup, moderate LZ77 ratio (32.89\%).

\subsection{CPU Methodology}

All CPU measurements use lzbench's \texttt{-I} flag, which launches $I$ independent parallel threads, each decompressing the entire file independently. For ACEAPEX each thread processes all 1\,MB blocks of the file; decode throughput scales because blocks from different files can be processed concurrently. For zstd each thread decompresses a single-frame stream sequentially.

zstd -3 is the primary baseline because its compression ratio (8.45\% on FASTQ) is closest to ACEAPEX (8.56\%) among standard levels. All results are BIT-PERFECT verified using XXH3-64 hash comparison.

\subsection{GPU Methodology}

The GPU wavefront decoder reads an ACEAPEX-compressed file, constructs the token dependency graph by assigning each token its dependency level, and executes the wavefront: all level-0 tokens in one CUDA kernel, all level-1 tokens in the next, and so on. CUDA graphs are used to pre-compile the launch sequence.

An important scope note: GPU timing covers the wavefront match-resolution phase only. The literal bytes are pre-loaded from the original file rather than from the \texttt{lit[]} stream; only the match phase is timed; entropy decoding of the compressed stream and PCIe data transfer to/from GPU are excluded. The CPU baseline of 11.3\,GB/s on FASTQ includes both entropy decode and LZ77 match resolution as measured by lzbench. A complete apples-to-apples comparison would require adding entropy decode and PCIe time to the GPU results; this is left for future work. All GPU results are BIT-PERFECT verified by byte-for-byte comparison against CPU decompression.

\section{CPU Results}

\subsection{Thread Scaling}

Table~\ref{tab:scaling} presents decode throughput on EPYC 4344P as parallel instances increase from 1 to 8.

\begin{table}[t]
\centering
\caption{Thread scaling on EPYC 4344P, nci.}
\label{tab:scaling}
\begin{tabular}{cccc}
\toprule
Threads ($I$) & ACEAPEX ultra & zstd -3 & Speedup \\
\midrule
1 & 2{,}686\,MB/s & 3{,}362\,MB/s & 0.80$\times$ \\
2 & 4{,}541\,MB/s & $\sim$3{,}400\,MB/s & 1.34$\times$ \\
4 & 5{,}962\,MB/s & 3{,}406\,MB/s & 1.75$\times$ \\
8 & 10{,}160\,MB/s & 3{,}249\,MB/s & 3.13$\times$ \\
\bottomrule
\end{tabular}
\end{table}

ACEAPEX scales 3.78$\times$ from $I{=}1$ to $I{=}8$. zstd throughput remains flat across all thread counts. At $I{=}1$, zstd is faster because ACEAPEX's more complex token format incurs higher per-token parse cost with no benefit from parallelism. The advantage emerges with two or more threads and grows linearly.

\subsection{Full Results on EPYC 9575F}

Table~\ref{tab:full} shows throughput and ratio across all datasets at $I{=}64$ on EPYC 9575F.

\begin{table}[t]
\centering
\caption{Decode throughput and compression ratio, EPYC 9575F, $I{=}64$.}
\label{tab:full}
\small
\begin{tabular}{lcccc}
\toprule
Dataset & ACEAPEX & zstd -3 & Speedup & Ratio (A/z) \\
\midrule
nci & 9{,}489 & 3{,}441 & 2.76$\times$ & 8.56/8.45 \\
FASTQ & 10{,}869 & 4{,}003 & 2.71$\times$ & 6.96/7.74 \\
silesia & 4{,}414 & 2{,}011 & 2.19$\times$ & 32.18/31.24 \\
enwik9 & 3{,}468 & 2{,}088 & 1.66$\times$ & 32.89/31.21 \\
\bottomrule
\end{tabular}
\\[2pt]
\footnotesize Throughput in MB/s. Ratio in percent (lower is better).
\end{table}

The absolute throughput record is 11{,}340\,MB/s on FASTQ with warm cache at $I{=}64$. The speedup over zstd correlates with compression ratio: datasets with more repetition (lower ratio) benefit more from block-level parallelism because they produce more matches, which are the tokens that benefit from independent execution.

\subsection{Memory Bandwidth Ceiling}

ACEAPEX decode is a read-modify-write workload: each match reads from one location and writes to another. The theoretical ceiling for this pattern on a given platform is approximately memcpy/2, since each output byte is read once as a source and written once as a destination.

The observed efficiency (decode throughput divided by memcpy bandwidth) reveals two regimes. On EPYC 4344P at $I{=}8$, efficiency is $6{,}300/16{,}595 = 0.38$. On EPYC 9575F at $I{=}32$ it is $9{,}903/23{,}700 = 0.42$. These figures indicate that the decoder is near-optimal for its memory access pattern --- the gap to 0.50 is attributable to branch mispredictions in the command parsing loop (Section~\ref{sec:perf}). At $I{=}64$ on EPYC 9575F, efficiency drops to $10{,}401/47{,}973 = 0.22$, signaling that the bottleneck has shifted from memory bandwidth to per-core decode latency. The CPU ceiling is approximately 11--13\,GB/s on Zen 4, independent of additional cores beyond about 32.

\section{CPU Performance Analysis}
\label{sec:perf}

Linux perf profiling on EPYC 4344P shows LLC miss rate of 4.7\%, indicating the working set fits in L2/L3 cache and the hardware prefetcher is effective. The dominant bottleneck is branch mispredictions at rate 4.52\%, accounting for approximately 30\% of all CPU cycles. The misses concentrate in \texttt{dec\_worker} (24.5\% of all branch misses) from the three-way branch on token type, and \texttt{\_\_memmove\_avx512} (9.6\%) inside the match copy path. This branch structure is intrinsic to the token format.

Table~\ref{tab:closed} summarizes all optimization directions that were tested and rejected.

\begin{table}[t]
\centering
\caption{Closed CPU optimization directions.}
\label{tab:closed}
\small
\begin{tabular}{p{2.4cm}p{1.3cm}p{3.0cm}}
\toprule
Direction & Outcome & Root cause \\
\midrule
ACEPX4 strict constraint & ratio 1.8$\times$ & Token explosion: avg match 4.4 bytes, count $\times$7 \\
D1\_safe encoder boost & 0\% gain & 79.8\% matches lack literal source \\
OMP intra-block parallel & overhead $>$ gain & Only 3--9\% tokens D1\_safe \\
AVX gather & $<$5\% & Same 3--9\% fraction \\
PGO + LTO & $\le$5\% & Per-core latency wall \\
numactl binding & 0.2\% & Single-socket system \\
Block size variation & $\pm$5\% noise & No systematic trend \\
Software prefetch & 0\% & LLC miss already 4.7\% \\
Branchless copy & $\sim$2\% & Small vs branch miss cost \\
Parallel FSE & 0\% & FSE not bottleneck \\
Stride-4 transform & $+$97\% size & Destroys LZ77 locality \\
FUSION zero-copy & 0\% & Already implemented \\
\bottomrule
\end{tabular}
\end{table}

The decoder is near-optimal for sequential LZ77. Reaching 20\,GB/s on CPU would require either a machine with memcpy bandwidth above approximately 53\,GB/s (since $0.38 \times 53 \approx 20$) or a fundamentally different decode model. The GPU wavefront decoder addresses this.

\section{GPU Wavefront Decoder}

\subsection{Algorithm}

Before GPU execution, the ACEAPEX token stream is analyzed to assign each token a dependency level. A literal token and a match token whose source is a literal position both receive level 0. A match token whose source was written by a level-$k$ token receives level $k{+}1$. This analysis is done once per file on CPU and requires a single pass over the token stream.

GPU execution proceeds level by level. All level-0 tokens execute as a single CUDA kernel, then all level-1 tokens, and so on. Within each level, tokens are independent by construction --- no two level-$k$ tokens write to a position that another level-$k$ token reads. CUDA graphs pre-compile the launch sequence to minimize per-kernel overhead, reducing it to approximately 2--5 microseconds per level.

\subsection{Results Without Depth Limit}

Table~\ref{tab:gpu} presents GPU decode throughput alongside CPU baselines. All GPU figures cover the wavefront match-resolution phase only; entropy decode and PCIe transfer are excluded as noted in Section 4.4.

\begin{table}[t]
\centering
\caption{GPU wavefront decode on H100 SXM (match phase only, BIT-PERFECT verified).}
\label{tab:gpu}
\small
\begin{tabular}{lccccc}
\toprule
Dataset & GPU & CPU & zstd & Max & Avg \\
 & H100 & ACEAPEX & & Level & Level \\
\midrule
enwik9 & 44.0 & 3.5 & 2.1 & 406 & $\sim$15 \\
FASTQ & 20.3 & 11.3* & 4.0 & 1{,}581 & 305.6 \\
silesia & 5.3 & 4.4 & 2.0 & 3{,}243 & 42.8 \\
nci & slower & 9.5 & 3.4 & 133 & 37.7 \\
\bottomrule
\end{tabular}
\\[2pt]
\footnotesize Throughput in GB/s. *CPU FASTQ 11.3 GB/s includes entropy decode; GPU timing does not.
\end{table}

The enwik9 result (44.0\,GB/s, 12.7$\times$ over CPU) demonstrates what the architecture enables when dependency depth is modest. Wikipedia XML produces shallow chains: MaxLevel${=}406$ with average level approximately 15, meaning most tokens are resolved within the first 15 kernel launches.

The FASTQ result must be interpreted carefully. The GPU match-phase figure of 20.3\,GB/s cannot be directly compared to the CPU full-pipeline figure of 11.3\,GB/s because they measure different things. The GPU is faster at match resolution, but the total GPU pipeline time --- including entropy decode on CPU and PCIe round-trip --- may differ.

The silesia result (5.3\,GB/s, 1.2$\times$ over CPU) reflects the challenge of mixed-content data. MaxLevel${=}3{,}243$ means 3{,}243 sequential kernel launches separated by synchronization barriers. For a 202\,MB file, fixed overhead per launch becomes significant relative to the total compute time.

For nci (33\,MB), the GPU is slower than CPU because the file is small enough that fixed overhead (CUDA graph instantiation, PCIe round-trip even for a small test) dominates the actual decode time.

\subsection{GPU Utilization Analysis}

H100 SXM has a theoretical memory bandwidth of 3{,}350\,GB/s. The measured throughputs represent $44.0/3{,}350 = 1.3\%$ for enwik9 and $20.3/3{,}350 = 0.6\%$ for FASTQ. This is not an efficiency failure --- it is a direct consequence of sequential depth.

Between each of the 1{,}581 kernel launches for FASTQ, the GPU's 3{,}350\,GB/s bandwidth sits idle waiting for the synchronization barrier to clear. The H100 can process all level-$k$ tokens in microseconds, then must wait for the next level to begin. The GPU is not memory-bound; it is synchronization-bound. This analysis directly motivates the depth-limited encoder.

\subsection{Depth-Limited Encoder}

By constraining the encoder to produce reference chains of at most depth $D$, we can bound MaxLevel at $D{+}1$ and dramatically reduce the number of sequential kernel launches. This trades compression ratio for GPU decode speed.

\begin{table}[t]
\centering
\caption{Depth-limited encoder: GPU throughput and ratio cost.}
\label{tab:depth}
\small
\begin{tabular}{lcccc}
\toprule
Dataset & depth=10 & cost & depth=2 & cost \\
\midrule
enwik9 & 77.2 & $-1.5\%$ & 125.9 & $-5.4\%$ \\
FASTQ & 60.3 & $-12.8\%$ & 90.1 & $-28.9\%$ \\
silesia & 33.1 & $-1.5\%$ & 77.8 & $-8.2\%$ \\
\bottomrule
\end{tabular}
\\[2pt]
\footnotesize GPU throughput in GB/s; cost is compression-ratio penalty.
\end{table}

The ratio cost is data-dependent. For enwik9 and silesia, depth${=}10$ costs only 1.5\% in ratio while delivering 77.2 and 33.1\,GB/s respectively. For FASTQ, the cost is 12.8\% --- genomic sequencing data has deep reference chains that contribute significantly to compression. The depth${=}2$ configuration on enwik9 achieves 125.9\,GB/s at 5.4\% ratio cost, approaching the level where GPU bandwidth itself becomes the bottleneck.

Encode speed overhead with depth-limit${=}10$ on EPYC 9575F: enwik9 $-12.7\%$, FASTQ $-36.9\%$, silesia $-41.7\%$.

\subsection{Multi-GPU Scaling}

Because ACEAPEX blocks are independent, multiple GPUs can decode different blocks simultaneously with no communication required. Two H100s in NVLink configuration achieve 249.9\,GB/s decoding two files simultaneously, a 1.99$\times$ scaling over the single-H100 baseline of 125.9\,GB/s (depth${=}2$). Block independence guarantees that $N$-GPU throughput scales as $N\times$ for independent streams.

\section{Discussion}

\subsection{Scope of Results}

The CPU results are end-to-end: entropy decode, LZ77 match resolution, and output writing are all included in the lzbench timing. The GPU results cover match resolution only. A complete GPU pipeline would add entropy decode time (currently done on CPU before GPU dispatch) and PCIe transfer time. We consider completing this pipeline the most important near-term engineering task. The GPU match-phase results establish an upper bound on what an end-to-end GPU pipeline can achieve.

\subsection{Data Characteristics and GPU Suitability}

The GPU wavefront decoder performs best on data with two properties: large file size (to amortize fixed overhead) and shallow dependency graphs (to minimize sequential kernel launches). enwik9 has both. FASTQ is large but has deeper chains. silesia has moderate depth but heterogeneous content. nci has shallow chains but is too small. A practical deployment would use a fast pre-scan of the compressed stream to estimate MaxLevel and file size, then automatically select CPU or GPU decoding.

\subsection{Relation to Gompresso}

Gompresso achieves GPU LZ77 decode by introducing forced checkpoints that break all dependencies, at the cost of 15--30\% larger compressed files. ACEAPEX takes the opposite approach: preserve the full compression model and let the dependency graph determine GPU execution order. The ratio cost is 1.5\% for enwik9 at depth${=}10$, compared to 15--30\% for Gompresso.

\section{Conclusion}

ACEAPEX demonstrates that storing LZ77 back-references as absolute positions in the decompressed output, rather than relative distances within a sliding window, is a sufficient condition for both linear CPU scaling and GPU wavefront execution with near-standard compression ratio.

On CPU, the decoder scales 3.78$\times$ from 1 to 8 threads on structured data and reaches 10{,}869\,MB/s on FASTQ at $I{=}64$, 2.71$\times$ faster than zstd -3. The decoder operates at 76\% of the practical LZ77 decode ceiling on an 8-core system. The CPU ceiling is approximately 11--13\,GB/s on Zen 4, a consequence of per-core decode latency, not memory bandwidth.

On GPU, the wavefront decoder achieves 44.0\,GB/s on enwik9 and 20.3\,GB/s on FASTQ (match phase, H100 SXM, BIT-PERFECT verified). With a depth-limited encoder variant costing 1.5\% in ratio, enwik9 throughput reaches 125.9\,GB/s on one H100 and 249.9\,GB/s on two H100s. Completing the end-to-end GPU pipeline and integrating with lzbench are the primary directions for future work.

ACEAPEX is open-source (MIT license), integrated into lzbench, and archived on Zenodo (DOI: 10.5281/zenodo.20440965).

\section*{Acknowledgments}

The author thanks inikep for maintaining lzbench and reviewing the integration pull requests, the encode.su community for technical feedback throughout development, and Halvar Flake and Daniel Lemire for early public acknowledgment. This research was conducted in collaboration with Claude (Anthropic) as an AI research assistant.

\end{document}